\documentclass[aps, prb, onecolumn, nofootinbib, citeautoscript, 10pt]{revtex4-2}

\usepackage[papersize={8.5in,11in}]{geometry}

\usepackage[colorlinks=true]{hyperref}
\hypersetup{
	bookmarks=true,         
	unicode=false,          
	pdftoolbar=true,        
	pdfmenubar=true,        
	pdffitwindow=false,     
	pdfstartview={FitH},    
	pdfkeywords={keyword1} {key2} {key3}, 
	pdfnewwindow=true,      
	colorlinks=true,       
	linkcolor=blue,           
	citecolor=blue,        
	filecolor=magenta,      
	urlcolor=blue           
} 

\usepackage{amsmath,amssymb, ulem, float} 
\usepackage{comment, cancel}
\usepackage[version=4]{mhchem}
\usepackage{dcolumn}
\usepackage[tight]{subfigure} 
\usepackage[dvipsnames]{xcolor} 
\usepackage{color, enumitem}
\usepackage{amssymb,amsmath}
\usepackage{tabularx, graphicx}
\usepackage{bbm,bm,array,physics}
\usepackage{dsfont}

\def \nn{\nonumber \\}

\begin{document}

\title{Puiseux series about exceptional singularities dictated by symmetry-allowed Hessenberg forms of perturbation matrices}

\author{Ipsita Mandal}
\email{ipsita.mandal@snu.edu.in}

\affiliation{Department of Physics, Shiv Nadar Institution of Eminence (SNIoE), Gautam Buddha Nagar, Uttar Pradesh 201314, India}


\begin{abstract}
We develop a systematic framework for determining the nature of exceptional points of $n^{\rm th}$ order (EP$_n$s) in non-Hermitian (NH) systems, represented by complex square matrices. By expressing symmetry-preserving perturbations ($\propto \epsilon$) in the Jordan-normal basis of the defective matrix at an EP$_n$, we show that the upper-$\ell$ Hessenberg structure of the perturbation matrix directly dictates the leading-order eigenvalue- and eigenvector-splitting to be $\propto  \epsilon^{1/\ell}$, when expanded in a Puiseux series. Applying this to three-band NH models invariant under parity (P), charge-conjugation (C), or parity-time-reversal (PT), we find that EP$_3$s in P- and C-symmetric systems are restricted to at most $\sim  \epsilon^{1/2}$ branch points, while PT-symmetric systems generically support EP$_3$s with the strongest possible singularities (viz. $\sim \epsilon^{1/3}$). We illustrate these results with concrete three-dimensional models in which exceptional curves and surfaces emerge. We further show that fine-tuned perturbations can suppress the leading-order branch point to a less-singular splitting, which have implications for designing direction-dependent EP-based sensors. The appendix extends the analysis to four-band P- and C-symmetric models, establishing the existence of EP$_4$s with $\sim \epsilon^{1/4}$ and  $\sim \epsilon^{1/2}$ singularities, respectively.
\end{abstract}

\maketitle

\tableofcontents

\section{Introduction}

Non-Hermitian (NH) topological phases have become a central topic in condensed matter physics, following the realisation that stable band-crossing points are more generic and abundant than those found in Hermitian systems \cite{emil_nh_nodal,yoshida-peters, emil_review,ips-emil,prl_eps}.
The robust topology of such stable NH phases \cite{emil_review} is characterised by invariance under various symmetries of complex matrices \cite{bernard, ips-emil}, strongly hinged on the mathematical concept of exceptional points (EPs). EPs are singular points in the parameter-space of complex matrices at which two or more eigenvalues coalesce, sharing a single linearly-independent eigenvector \cite{BerryDeg, Heiss, PhysRevX.6.021007, EPreview, ozdemir2019parity}. EPs can be understood as the natural dissipative analogues of the stable nodal points \cite{BerryDeg, Heiss} familiar from the Weyl nodes, which are the well-known Hermitian systems serving as the poster-child of the concept of topology realised in solid-state systems. The emergence of EPs, in fact, give rise to a rich array of topologically-robust phenomena in generic NH scenarios, examples including bulk Fermi arcs \cite{NHarc} and unidirectional lasing \cite{lasing}. All these have propelled the understanding of EPs to an interdisciplinary research frontier, with their relevance spanning classical metamaterials and quantum condensed matter systems \cite{EPreview}. In this context, it is worthwhile to mention that the intimate connection of EPs with topological phase transitions in diverse contexts has been firmly established in our current scientific knowledge, as they are invoked in widely-different studies \cite{ips-ep-epl, ips-tewari, emil_review, kang-emil, ips-emil, prl_eps,  ips-kang, ips-bp, ips-yao-lee}.

While the initial studies involving spectral singularities of \textit{NH Hamiltonians} were directed toward second-order EPs (EP$_2$s), which arise naturally in two-dimensional (2D) NH systems \cite{koziifu, NHarc}, the abundance of higher-order EPs (e.g., third-order EPs) has been demonstrated in Ref.~\cite{ips-emil}. Exceptional curves and surfaces of second- and higher-order have also been identified under relevant circumstances \cite{EPrings, emil-johan, EPringExp, emil-marcus}. From a general perspective, realising an $n$-th order EP (which we will label as EP$_n$) requires the tuning of $(2n-2)$ real parameters, such as components of the lattice-momentum vector. This counting argument implies that EPs of order $n \geq 3$ can only be expected in at least four-dimensional systems. Nevertheless, we have established that generic NH symmetries can stabilise EP$_3$s in 2D systems. In particular, this implies that three-dimensional (3D) systems will host such exceptional singularities in the form of nodal curves, which we will call exceptional curves (ECs). In this paper, we continue the ongoing efforts to characterise the endless avatars of exceptional singularities by digging into how the degree of the singularity originates from the overall structure of the complex matrix. Since the EPs are associated with branch-point singularities, an expansion about an EP would be dictated by a Puiseux series \cite{kato, ma-pert, Demange_2012}, which is  a generalised power series allowing for negative and fractional exponents of the indeterminate. In particular, we study three-band NH models, which are represented by $3\times 3$ complex matrices invariant under parity (P), charge-conjugation (C), or parity-time-reversal (PT).

The paper is organised as follows: In Sec.~\ref{sechess}, we explain the mathematical objects and concepts needed to understand the nature of possible perturbations to an $s \times s$ complex matrix, which can be taken to correspond to an $s$-band NH model in the context of solid-state systems. In Sec.~\ref{secep2}, we review and apply the idea of perturbations in the form of Hessenberg matrices to a $2\times 2$, which produces the simplest-possible singular degeneracies, namely EP$_2$s. Sec~\ref{secep3} is devoted to deriving the generic forms of the singularities for three-band models under the protection of 3 different kinds of symmetries, spanning P-, C-, and PT-symmetries. There, we supplement our analysis of generic matrices with specific 3D systems in the momentum space. We end with some discussions and outlook in Sec.~\ref{secsum}.
The appendix deals with analysing C- and P-symmetric 4-band models.
 
\section{Perturbation around an exceptional degeneracy}
\label{sechess}

In this section, we provide the mathematical definitions required to understand the singularity structure of exceptional degeneracies for generic complex matrices. An $s\times s$ matrix which has $(\ell-1)$ sub-diagonals, adjacent to the principal diagonal in the lower half, which do not identically comprise zero elements, is referred to as a upper-$\ell$ Hessenberg matrix, $B_{s, \ell}$. In other words, if the matrix-elements are denoted as $m_{i,j}$, then we have $m_{i, j} = 0 $ for $i > j + \ell-1$. In this language, the common nomenclature of an upper Hessenberg matrix is actually an upper-2 Hessenberg matrix. Let us consider a Jordan block, $ \mathcal {J}_s (\lambda) $,
of dimension $s$ and $s$-fold degenerate eigenvalue, $\lambda $. By examining the perturbed matrix, $ \mathcal {J}_s + \epsilon \, B_{s, \ell} $, Ref.~\cite{ma-pert} has derived how the degenerate eigenvalues of $ \mathcal {J}_s  $ split under a perturbation of the form  $\epsilon \, B_{s, \ell}$. The main result can be summarised as the nature of the eigenvalues ($\lbrace \lambda_{i, n_{s_i}} \rbrace $) of all possible perturbed matrices, can be grouped in the following manner, characterised by the respective Puiseux series:
\begin{enumerate}

\item Prepare all possible sets of $\lbrace s_i \rbrace $, such that $\sum_i s_i =  s \,.$

\item For each $i$, we have a ring of size $s_i$, defined by the elements,
\begin{align}
\lambda_{i, n_{s_i}} = \lambda + \alpha_{i, 1}\, \omega_{s_i}^{n_{s_i}} \,\epsilon^{\frac{1}{s_i}}
+ \alpha_{i, 2}\, \omega_{s_i}^{ 2\,n_{s_i}} \,\epsilon^{\frac{2}{s_i}} + \cdots\,,
\quad \omega_{s_i} = e^{\frac{2\,\pi\,i} {s_i}}\,,
\text{ for } n_{s_i} \in [1, s_i]\,.
\end{align}

\end{enumerate}

Let us illustrate this for some values of $s$.
\begin{enumerate}

\item \underline{$s = 2$:}
\\We can have two sets, viz. $\lbrace s_1=1, \, s_2 = 1 \rbrace$ and $ \lbrace s_1= 2 \rbrace $. For set 1, we have $\omega_{s_1} = \omega_{s_2} = \omega_1 \equiv 1$ and the eigenvalues behaving as
\begin{align}
\lambda_{1, 1} = \lambda + \alpha_{1, 1}\, \epsilon
+ \alpha_{1, 2}\,\epsilon^2 + \cdots \text{ and }
\lambda_{2, 1} = \lambda + \alpha_{2, 1}\, \epsilon
+ \alpha_{2, 2}\,\epsilon^2 + \cdots \,.
\end{align}
These give rise to two rings of the trivial size of unity.
For set 2, we have $\omega_{s_1} =  \omega_2 \equiv -1 $ and the eigenvalues behaving as
\begin{align}
\lambda_{1, 1} = \lambda - \alpha_{1, 1}\,\epsilon^{\frac{1}{2}}
+ \alpha_{1, 2}\, \epsilon + \cdots \text{ and }
\lambda_{1, 2} = \lambda + \alpha_{1, 1} \,  \epsilon
+ \alpha_{1, 2} \, \epsilon^2 + \cdots \,.
\end{align}
These constitute one ring of size 2.
 
\item \underline{$s = 3$:}
\\We can have three sets, viz. $\lbrace s_1=1, \, s_2 = 1, \, s_3 = 1 \rbrace$, $ \lbrace s_1= 2 , \, s_2 = 1 \rbrace $,
and $ \lbrace s_1= 3 \rbrace $.
\\For set 1, we have $\omega_{s_1} = \omega_{s_2} = \omega_{s_3} = \omega_1 \equiv 1$ and the eigenvalues behaving as
\begin{align}
\lambda_{1, 1} = \lambda + \alpha_{1, 1}\, \epsilon
+ \alpha_{1, 2}\,\epsilon^2 + \cdots\,, \quad
\lambda_{2, 1} = \lambda + \alpha_{2, 1}\, \epsilon
+ \alpha_{2, 2}\,\epsilon^2 + \cdots \,, \text{ and }
\lambda_{3, 1} = \lambda + \alpha_{3, 1}\, \epsilon
+ \alpha_{3, 2}\,\epsilon^2 + \cdots \,.
\end{align}
These give rise to three rings of the trivial size of unity.
For set 2, we have $\omega_{s_1} =  \omega_2 \equiv - 1 $, $\omega_{s_2} =  \omega_1 \equiv 1 $, and the eigenvalues behaving as
\begin{align}
\label{ep3-sqrt}
\lambda_{1, 1} = \lambda - \alpha_{1, 1}\, \epsilon^{\frac{1}{2}}
+ \alpha_{1, 2}\, \epsilon + \cdots\,, \quad
\lambda_{1, 2} = \lambda + \alpha_{1, 1} \, \epsilon^{\frac{1}{2}}
+ \alpha_{1, 2} \,\epsilon + \cdots \,, \text{ and }
\lambda_{2, 1} = \lambda + \alpha_{2, 1}\, \epsilon
+ \alpha_{2, 2}\,\epsilon^2 + \cdots \,.
\end{align}
These constitute one ring of size 2 and one trivial ring of size unity.
For set 3, we have $\omega_{s_3} =  \omega_3  \equiv e^{\frac{2\,\pi\,i} {3}} $ and the eigenvalues behaving as
\begin{align}
& \lambda_{1, 1} = \lambda + \alpha_{1, 1}\, \omega_3 \,\epsilon^{\frac{1}{3}}
+ \alpha_{1, 2}\, \omega_3^2 \,\epsilon^{\frac{2}{3}} 
+ \alpha_{1, 3}\, \omega_3^3 \,\epsilon + \cdots\,, \quad
\lambda_{1, 2} = \lambda + \alpha_{1, 1} \, \omega_3^2\, \epsilon^{\frac{1}{3}}
+ \alpha_{1, 2} \, \omega_3^4 \,\epsilon^{\frac{2}{3}} 
+ \alpha_{1, 3}\, \omega_3^6 \,\epsilon + \cdots \,,
\nn & \text{and }
\lambda_{1, 3} = \lambda + \alpha_{1, 1} \, \omega_3^3\, \epsilon^{\frac{1}{3}}
+ \alpha_{1, 2} \, \omega_3^6 \,\epsilon^{\frac{2}{3}} 
+ \alpha_{1, 3}\, \omega_3^9 \,\epsilon + \cdots\,.
\end{align}
These constitute one ring of size 3.

\end{enumerate}
We emphasise that, for $\epsilon \neq 0$, the perturbed matrix is generically nondegenerate and, hence, is no longer an EP in the strict sense of exact degeneracy plus defectiveness --- the terminology `singular' refers instead to $\epsilon=0$ remaining a branch point of a multivalued function, whose Puiseux expansion starts at the fractional power $\epsilon^{1/\ell}$ rather than at $\epsilon^1$.

\section{Review of exceptional points in two-band models}
\label{secep2}

For two-band models, we can get only EPs of the lowest-possible order, viz. EP$_2$, because a perturbation in the form of a Hessenberg matrix can only be of the upper-2 type. Nevertheless, we demonstrate this point explicitly by taking an example. Consider the non-Hermitian matrix
\begin{align}
\mathcal H_2 = \begin{bmatrix}
0 & a\\
1 & 0 \\
\end{bmatrix}.
\end{align}
For $a=0$, it reduces to a singular matrix hosting an EP$_2$ with degenerate eigenvalues $\lambda = 0 $. Its Jordan-normal form is
\begin{align}
S^{-1} \, \mathcal H_2  \, S = \mathcal J_2 (0)\,, \text{ with }
S = \begin{bmatrix}
0 & 1\\
1 & 0 \\
\end{bmatrix}.
\end{align}
Adding a perturbation $\epsilon \, B_{2,2}$, with
\begin{align}
B_{2,2} = \begin{bmatrix}
0 & 0\\
\epsilon  & 0 \\
\end{bmatrix}, \text{ where }
S\, B_{2,2}\, S^{-1} = \begin{bmatrix}
0 & \epsilon \\
0  & 0 \\
\end{bmatrix},
\end{align}
the eigenvalues of the perturbed matrix, $\mathcal H_2 + \epsilon\,  S\, B_{2,2}\, S^{-1}$ (or, equivalently, $\mathcal J_2 + \epsilon\, B_{2,2} $) are given by $ \pm \sqrt \epsilon$. Thus the generic behaviour around an EP$_2$ is that around a square-root branch point.

\section{Exceptional singularities in three-band models}
\label{secep3}

In this section, our aim is to explore the emergence of higher-order EPs in three-band models, because they are able to host EP$_3$s \cite{ips-emil}. We show how symmetry-restrictions dictate the leading-order singular nature of these EPs, focusing on P-, C-, and PT-symmetry classes \cite{bernard}. Depending on the symmetry class, the EPs might show cube-root or square-root branch points. We also provide some examples, falling into one of these classes, obtained from NH generalisations of 3D bandstructures.

\subsection{P-symmetric systems}

The imposition of the P-symmetry to a $3\times 3$ non-Hermitian matrix leads to the form \cite{ips-emil},
\begin{align}
\mathcal H^{\rm P}_3 = \begin{bmatrix}
0 & b & c \\
 d & 0 & 0 \\
 g & 0 & 0 \\
\end{bmatrix}, \quad
P = \text{diag}\lbrace -1,\,1,\,1 \rbrace ,\quad
P\,\mathcal H^{\rm P}_3\, P^{-1} = - \, \mathcal H^{\rm P}_3 \,.
\end{align}
The $3\times 3$ matrix $P$ forms a three-dimensional realisation of the P-symmetry. We note that, an unbroken P symmetry constrains the set of eigenvalues, $ \lbrace E_i \rbrace$, to obey the condition $\lbrace E_i \rbrace = \lbrace - E_j \rbrace $ \cite{emil_nh_nodal}.
Consequently, for an odd-dimensional square matrix, there is always one zero eigenvalue (i.e., flat or non-dispersive band). For $b = -\,{c\, g} / {d}$, it hosts an EP$_3$ with the threefold-degenerate eigenvalue, $\lambda = 0$. At that point, its Jordan-normal form is
\begin{align}
S^{-1} \, {\mathcal H^{\rm P}_{\rm EP}} \, S = \mathcal J_3(0)\,, \text{ with }
S = \begin{bmatrix}
 0 & \frac{1}{g} & 0 \\
 \frac{d}{g} & 0 & -\frac{d}{c \,g^2} \\
 1 & 0 & 0 \\
\end{bmatrix} 
\text{ and }
{\mathcal H^{\rm P}_{\rm EP}}  = {\mathcal H^{\rm P}_3} \Big \vert_{b =\frac{- c\, g}  {d}}.
\end{align}
Adding a perturbation $\epsilon\, \tilde B^{\rm P}_{3,2}$ to ${\mathcal H^{\rm P}_{\rm EP}}$, with
\begin{align}
\tilde B^{\rm P}_{3,2} = \begin{bmatrix}
0 & a_1 & a_2 \\
 a_3 & 0 & 0 \\
 a_4 & 0 & 0 \\
\end{bmatrix},  
\text{ where }
S^{-1}\, \tilde B^{\rm P}_{3,2}\, S = \begin{bmatrix}
 0 & \frac{a_4}{g} & 0 \\
 a_1\, d+a_2\, g & 0 & -\frac{a_1 \,d}{c\, g} \\
 0 & c \left(a_4-\frac{a_3 \,g}{d}\right) & 0 \\
 \end{bmatrix} .
\end{align}
the three eigenvalues are obtained as $\left\{0, \, \pm \,\varepsilon_2 \right\}$, where $ \varepsilon_2 =
 \sqrt{c \left(a_4-\frac{a_3 g}{d}\right)+a_1 \left(a_3 \epsilon +d\right)+a_2 \left(a_4 \epsilon +g\right)}\; \epsilon^{1/2} $. This conforms to the generic form shown in Eq.~\eqref{ep3-sqrt}. We note that a P-symmetry-preserving perturbation takes the form of an upper-2 Hessenberg matrix, $B_{3,2}$, in the Jordan-normal basis of the defective matrix, ${\mathcal H^{\rm P}_{\rm EP}} $. And as commented above, the flat-band remains unaffected as long the P-symmetry remains unbroken. In particular, an upper-3 Hessenberg matrix, $B_{3,3}$, would break the P-symmetric form, although it could produce a leading-order splitting $\sim \epsilon^{1/3}$. One more crucial observation is that, on setting $ a_4 =  {a_3\, g}/{d}
-(a_1\, d + a_2 \,g) / {c} $ , the leading-order behaviour of $\varepsilon_2$ reduces to $\sim \epsilon $, since the coefficient of the square-root term vanishes.
The directional dependence discussed here should not be interpreted as a change in the nature of the underlying EP. The EP$_3$ remains intact, since the defective Hamiltonian at $\epsilon=0$ continues to possess a single $\mathcal{J}_3(0)$ Jordan block. Tuning parameters, such as $a_4$, we can select a special symmetry-preserving perturbation direction for which the coefficient of the leading $\epsilon^{1/2}$ term in the Puiseux expansion vanishes. The splitting is then governed by the next non-vanishing contribution, resulting in a linear dependence on $\epsilon$. This does not correspond to a reduction to an EP$_2$ or to a Dirac point; instead, it represents a direction-dependent linear dispersion away from the same EP$_3$. The fine-tuned directions form a codimension-one subset within the symmetry-allowed perturbation space, while generic directions recover the usual $\epsilon^{1/2}$ behaviour.

Let us now consider a physical example in the form of the continuum model for a node carrying a dipole of Berry curvature (BC), which constitutes a three-band Hopf semimetal \cite{graf-hopf, ips-vnr, ips-fa, ips-fa-lett}. A simple generalised NH version can be written as:
\begin{align}
\label{eqhopf}
\mathcal H^{\rm hopf}_3 = \begin{bmatrix}
0 & k_x-i \, k_y & -i\,k_z + i\, a\\
k_x-i \, k_y & 0 & 0 \\ 
i\,k_z+i\, a  & 0 & 0 \\
\end{bmatrix},
\end{align}
obeying $P\,\mathcal H^{\rm hopf}_3\, P^{-1} = - \, \mathcal H^{\rm hopf}_3 $.
The non-Hermiticity appears for a nonzero value of $a$, which we assume to be real here. The system has a continuous surface of exceptional singularities of order 3 (which we label as ES$_3$) on the sphere, $k_x^2 + k_y^2 + k_z^2 = a ^2$. The ES$_3$ is characterised by the Jordan-normal form, $ \mathcal {J}_3 (0) $. Let the ES$_3$ be parametrised by $k_x = a \sin \theta \cos \phi $, $k_y = a \sin \theta \sin \phi $, and $ k_z = a \cos \theta $. A P-symmetry-preserving perturbation of the form of $\epsilon \, \tilde B^{\rm P}_{3,2} $, with respect to the singular matrix at the location of the ES$_3$, produces the eigenvalues,
$$\lbrace 0,\, \pm  \left[
a \, e^{-i  \,\alpha } a_3 \sin \theta + a_1 \left(a_3 \, \epsilon + a \,e^{i\, \alpha } \sin \theta \right)-i \,a\, a_4
   (\cos \theta -1)+a_2 \left \lbrace a_4 \,\epsilon +i \,a (\cos \theta +1)\right \rbrace
   \right ] \;\epsilon^{1/2}\rbrace .$$ 
If we set $a_4 = \frac{a_2 (\cos \theta + 1)-i \,e^{-i \,\alpha } \left(a_3+e^{2\, i \, \alpha } a_1\right) \sin \theta }
{\cos \theta -1}$, the nonzero eigenvalues behave as $ \sim  \epsilon $, showing that the coefficient of the $\epsilon^{1/2}$ term is adjusted to vanish for this choice of parameters.

\begin{figure*}[t!]
\centering
\subfigure[]{\includegraphics[width= 0.35 \textwidth]{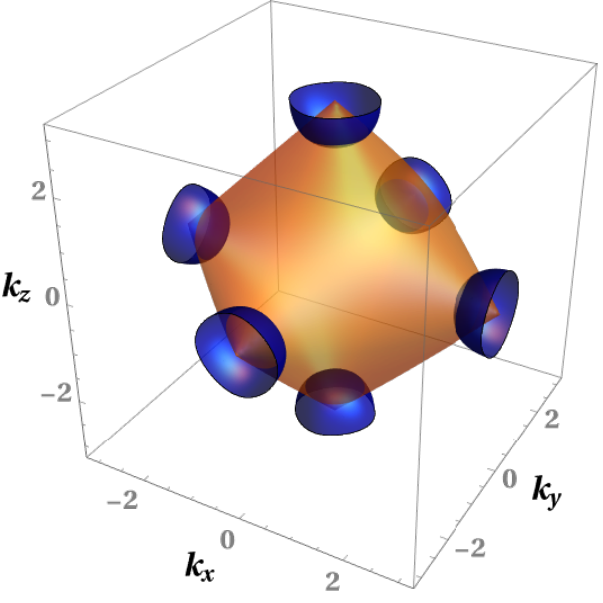}} \hspace{2 cm}
\subfigure[]{\includegraphics[width= 0.35 \textwidth]{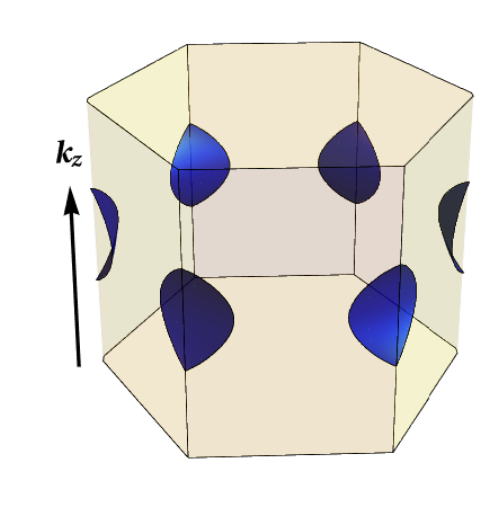}}
\caption{Third-order exceptional degeneracies for the lattice models represented by Eq.~\eqref{eqhopf-lat}. (a) 3 pairs of EC$_3$ curves show up on setting $\Delta =-1$ and $a =0.75 $ in $\mathcal H^{\rm hopf1}_3$. (b) 3 pairs of exceptional surfaces show up on setting $a = 1 $ in $\mathcal H^{\rm hopf2}_3$.\label{figepadipole}}
\end{figure*}
 
Two parent lattice versions of the continuum model of Eq.~\eqref{eqhopf} are given by
 \begin{align}
\label{eqhopf-lat}
\mathcal H^{\rm hopf1}_3 &= \begin{bmatrix}
0 & \sin k_x-i \sin k_y & \Delta + e^{-i\, k_z}+\cos k_x +\cos k_y + i\, a\\
\sin k_x-i \sin k_y & 0 & 0 \\ 
\Delta + e^{i\, k_z}+\cos k_x +\cos k_y + i\, a  & 0 & 0 \\
\end{bmatrix} \nn 
\text{and }\mathcal H^{\rm hopf2}_3 & = \begin{bmatrix}
0 & w_k & -i\sin k_z + \frac{2\,i\, a} {3}\\
w_k^* & 0 & 0 \\ 
i \sin k_z+ \frac{2\,i\, a} {3}  & 0 & 0 \\
\end{bmatrix}, \nn\text{ where }
w_k  &= \frac{2}{3} \sum_{j=1}^3 e^{ i\, {\boldsymbol k}\cdot {\boldsymbol \delta}_j },\quad
{\boldsymbol \delta}_1 =\frac{1}{2}(\sqrt 3,\, 1,\,0) \,,\quad
{\boldsymbol \delta}_2 =\frac{1}{2}( -\,\sqrt 3,\, 1,\,0)\,,\quad
{\boldsymbol \delta}_3 = (0,\,- 1,\,0)\,.
\end{align}
The eigenvalues are: $\lbrace 0, \, \pm \sqrt{f_1+ i\, f_2} \rbrace $ and $\lbrace 0, \, \pm \sqrt{ f_3}\rbrace $ for the 2 systems, respectively, where
\begin{align}
f_1 &= 3+\Delta ^2+2 \cos k_x \left(\Delta +\cos k_y +\cos k_z \right) + 2 \, \Delta  \left(\cos k_y +\cos k_z \right)
+ 2 \cos k_y \cos k_z -a^2 \,,\nn
f_2 &= 2 \,a \left(\Delta +\cos  k_x +\cos  k_y +\cos k_z\right),\nn
f_3 & = \frac{4}{9}\left [3+4 \cos \left(\frac{\sqrt{3}\, k_x}{2}\right) \cos \left(\frac{3\, k_y}{2}\right)
+2 \cos \left(\sqrt{3} \, k_x\right) - a^2 \right ] +  \sin^2 k_z\,.
\end{align}

For the original Hermitian version with $a=0$, $\mathcal H^{\rm hopf1}_3$ corresponds to systems harbouring an odd number of nodal points harbouring BC-dipoles in the Brillouin zone (BZ), on setting $\Delta $ to $\pm 1$ or $\pm 3$ \cite{graf-hopf}. For the case of $\Delta = -1$, these nodal points appear at the points $ (\pi, \,0,\, 0)$, $ (0,\,\pi, \,0 )$, and $ (0,\, 0,\,\pi)$ in the BZ. The effect of adding a nonzero $a$ is that each nodal point transfigures into an exceptional curve of third-order (let us call it EC$_3$), which is the intersection of the two surfaces, $f_1 = 0 $ and $f_2=0$. An example of this behaviour is illustrated in Fig.~\ref{figepadipole}(a), where we have set $\Delta = 1$ and $ a=0.75 $ --- we get 3 pairs of EC$_3$ curves.

For the original Hermitian version with $a=0$, $\mathcal H^{\rm hopf2}_3$ corresponds to systems harbouring an even number of nodal points harbouring BC-dipoles in a tetragonal BZ, located at $k_z = 0$ \cite{graf-hopf}. Because the Bravais period gets doubled
along the $\boldsymbol{\hat z}$-direction, the $k_z$-values run from $-\pi/2$ to $\pi/2$ (i.e., its range gets halved). Thus a node can appear only for $k_z = 0$. The $k_x k_y$-plane at each $k_z$ forms a 2D slice in the shape of a hexagonal BZ (hBZ) and, basically, the dispersion is identical to the dispersion of graphene when Dirac points emerge at the $K$ and $K^\prime$ points of the hBZ. If we take the entire 3D picture, there are inequivalent BC-dipoles at $K$ and $K^\prime$, because the dipole-axes' orientations are opposite (i.e., along $ \pm \boldsymbol{\hat k }_z$).
Upon adding a nonzero $a$, each dipole transfigures into an ES when $\sqrt{ f_3}$ transitions from purely real to purely imaginary --- thus a regular nodal-point degeneracy blows up into a threefold-degenerate eigenvalue surface, displaying square-root singularity. We illustrate this via an example in Fig.~\ref{figepadipole}(b) by choosing $ a= 1 $ --- we observe three pairs of ES$_3$s.

\subsection{C-symmetric systems}

\begin{figure*}[t!]
\centering
\includegraphics[width= 0.25 \textwidth]{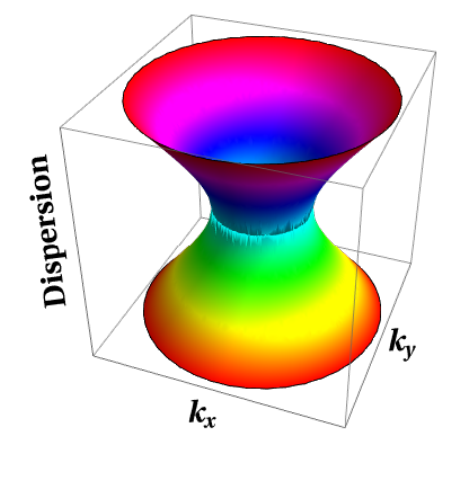}
\caption{A third-order exceptional circle (EC$_3$) for a three-band model, exemplified by Eq. \eqref{eqtsm}, showing square-root singularities.\label{bc-ep3}}
\end{figure*}

Let us consider a C-symmetric non-Hermitian matrix,
\begin{align}
\label{eqcsym}
\mathcal H^{\rm C}_3 = \begin{bmatrix}
 d & b & 0 \\
 c & 0 & b \\
 0 & c & -d \\
\end{bmatrix}, \quad
C = \begin{bmatrix}
 0 & 0 & 1 \\
 0 & -1 & 0 \\
 1 & 0 & 0 \\
\end{bmatrix},\quad
C \left ( \mathcal H^{\rm C}_3\right )^T C^{-1} = \epsilon_c \, \mathcal H^{\rm C}_3  \text{ with } \epsilon_c=-1 \,.
\end{align}
Here, $\lbrace b, \, c,\, d \rbrace $ are independent complex numbers. Just like the P-symmetric case, the C-symmetry with $\epsilon_c = -1 $ constrains the set of eigenvalues, $ \lbrace E_i \rbrace$, to obey the condition $\lbrace E_i \rbrace = \lbrace - E_j \rbrace $ \cite{emil_nh_nodal}. This implies that there will always be one zero eigenvalue (i.e., flat or non-dispersive band) for three-band systems.
For $ d = \pm\,i\,\sqrt{2\, b\, c} $, it hosts an EP$_3$ with a threefold-degenerate eigenvalue, $\lambda = 0$. Picking the value $ d = i\,\sqrt{2\, b\, c} $, the Jordan-normal form at this EP$_3$ is
\begin{align}
S^{-1} \, {\mathcal H^{\rm C}_{\rm EP}}  \, S = \mathcal J_3(0)\,, \text{ with }
S = \begin{bmatrix}
-\frac{b}{c} & \frac{i \, \sqrt{2\, b\, c}}{c^2} & \frac{1}{c^2} \\
 \frac{i\, \sqrt{2 \,b}} {\sqrt c} & \frac{1}{c} & 0 \\
 1 & 0 & 0 \\
\end{bmatrix} \text{ and }
{\mathcal H^{\rm C}_{\rm EP}}  = {\mathcal H^{\rm C}_3} \Big \vert_{ d = i\,\sqrt{2\, b\, c}}.
\end{align}
We add a C-symmetry-preserving perturbation $\epsilon\, \tilde B^{\rm C}_{3,2}$ to ${\mathcal H^{\rm C}_{\rm EP}} $, with
\begin{align}
\tilde B^{\rm C}_{3,2} = \begin{bmatrix}
a_1 & a_2 & 0 \\
 a_3 & 0 & a_2 \\
 0 & a_3 & -a_1 \\
\end{bmatrix},  \text{ where }
S^{-1}\, \tilde B^{\rm C}_{3,2}\, S = 
\begin{bmatrix}
 -a_1+\frac{i \sqrt{2 \,b}\; a_3 } {\sqrt{c}} & \frac{a_3}{c} & 0 \\
 i \,\sqrt{2 \,b \,c}\; a_1 + a_3 \,b+a_2\, c & 0 & \frac{a_3}{c} \\
 0 & i\, \sqrt{2 \,b \,c}\, a_1 + a_3 \,b + a_2 \,c & a_1-\frac{i \,\sqrt{2\, b} \;a_3 } {\sqrt{c}} \\
 \end{bmatrix}.
\end{align}
The three eigenvalues are obtained as $\left\{0, \, \pm \,\varepsilon_2 \right\}$, where $ \varepsilon_2 =
\sqrt{2\, i\, \sqrt{2\, b\, c} \; a_1 + 2 \, a_3\, b + 2 \, a_2 \left(a_3\, \epsilon +c\right) + a_1^2 \,\epsilon }
\; \epsilon^{1/2} $. This conforms to the generic form shown in Eq.~\eqref{ep3-sqrt}.
We note that a C-symmetry-preserving perturbation takes the form of an upper-2 Hessenberg matrix, $B_{3,2}$, in the Jordan-normal basis of the defective matrix, ${\mathcal H^{\rm C}_{\rm EP}} $. In particular, an upper-3 Hessenberg matrix, $B_{3,3}$, would break the C-symmetric form, although it could produce a leading-order splitting $\sim \epsilon^{1/3}$. One more crucial observation is that, on setting $ a_1 =  {i\, \left(a_3\, b
+ a_2 \, c\right)} /{\sqrt{2\, b\, c}} $ , the leading-order behaviour of $\varepsilon_2$ reduces to $\sim \epsilon $, since the coefficient of the square-root term vanishes.

Let us now demonstrate the consequences using a specific example in the form of the continuum model for a pseudospin-1 node \cite{ips-spin1-ph, ips-internode, ips-exact-spin1}. A simple generalised NH version can be written as:
\begin{align}
\label{eqtsm}
\mathcal H^{\rm sp1}_3 = \begin{bmatrix}
 k_z+ i \,a & \frac{k_x-i \,k_y}{\sqrt{2}} & 0 \\
 \frac{k_x + i \,k_y}{\sqrt{2}} & 0 & \frac{k_x-i \,k_y}{\sqrt{2}} \\
 0 & \frac{k_x+i\, k_y}{\sqrt{2}} & -k_z-i \,a \\
\end{bmatrix},
\end{align}
obeying $C \left[\mathcal H^{\rm spin1}_3\right]^T\, C^{-1} = - \, \mathcal H^{\rm spin1}_3 $ for the realisation of the symmetry shown in Eq.~\eqref{eqcsym}. The non-Hermiticity appears for a nonzero value of $a$, which we take to be real here, with eigenvalues equalling $\lbrace
0, \, \pm \,\sqrt{k_x^2+k_y^2+k_z^2- a^2 + 2 \,i\, a\, k_z } \rbrace $. The system hosts an EP$_3$ when $ k_x = a \cos \alpha $, $ k_y = a \sin \alpha $, and $ k_z = 0 $, characterised by the threefold-degenerate eigenvalue of zero and a single linearly-independent eigenvector $\propto \left [  -1, \,i \,\sqrt{2}\, e^{i \,\alpha }, \,e^{2\, i\, \alpha } \right]^T $. 
The singularity is actually an EC$_3$, as illustrated in Fig.~\ref{bc-ep3}, and the Jordan-normal form the defective matrix is given by $ \mathcal {J}_3 (0) $. A C-symmetry-preserving perturbation of the form of $ \epsilon\, \tilde B_{3,2}$, with respect to the location of the EC$_3$, produces the eigenvalues, $\lbrace 0,\, \pm \,
\sqrt{ \sqrt{2} \,  a  \, e^{-i  \, \alpha } \left(a_3 + e^{2  \, i  \, \alpha } \,  a_2\right)+\left(a_1^2 + 2 \, a_2 \,  a_3\right) \epsilon +2 \,  i  \, a \,  a_1} \; \epsilon^{1/2} \rbrace $.

\subsection{PT-symmetric systems}


In Ref.~\cite{ips-emil} (see also the supplemental material therein), we have derived the generic form of a three-band NH matrix obeying PT-symmetry. It has the form of
\begin{align}
H_3^{\text{PT}}  = \begin{bmatrix} 
 real_1 & i \,real_2 & real_3 \left(1 - i \right ) \\
i\, real_4 & real_5 & real_6  \left(1 + i \right ) \\
 real_7 \left(1+i \right ) & real_8 \left(1-i \right ) & real_9 
   \end{bmatrix},
\end{align}
which obeys the constraint $ H_3^{\text{PT}} = \left( P\, K \right)
 \left(H_3^{\text{PT}} \right)^* \left( P\, K \right)^{-1} $, with $ P =\text{diag}\lbrace 1,\,-1,\,1 \rbrace $ and $ K =\text{diag}\lbrace 1,\,1,\,i \rbrace $. As conveyed by the nomenclature, the 9 independent parameters, $\lbrace real_1, real_2, real_3, real_4, real_5, real_6, real_7, real_8, real_9 \rbrace $, are all real. The characteristic polynomial is of cubic order, which can be used to find the eigenvalues using the Cardano's method.

\begin{figure*}[t]
\centering
\subfigure[]{\includegraphics[width= 0.35 \textwidth]{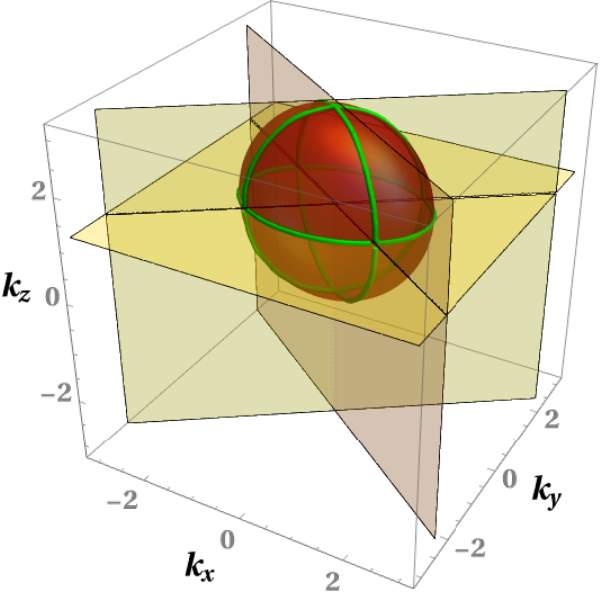}} \hspace{2 cm}
\subfigure[]{\includegraphics[width= 0.35 \textwidth]{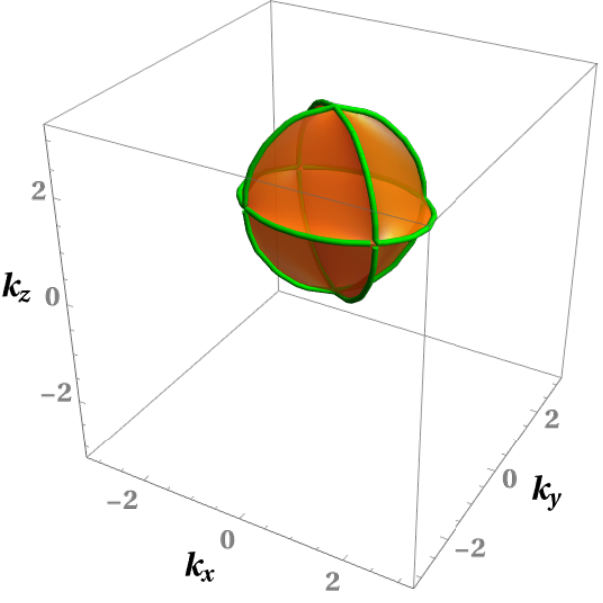}}
\caption{Using $a=1.25$ in Eq.~\eqref{eqPTev}: (a) The surfaces arising from $p=0$ (red) and $q=0$ (light yellow), which intersect to give a knotted curve (green) representing EC$_3$. (b) The surface of ES$_2$ representing $p^3 + q^2 = 0 $ (orange), on which the curve of EC$_3$ (green) is superimposed.\label{figpt}}
\end{figure*}

A specific linear-in-momentum model, which conforms to the above PT-symmetric form and exhibits EC$_2$s and EP$_3$s, has been provided in Ref.~\cite{ips-emil}. Here, we work with a 3d version, captured by
\begin{align}
\label{eqpt}
\mathcal H^{\rm PT}_3= \begin{bmatrix}
k_z-a & i\, k_x & (-1+i) a \\
 -i \,k_x & 0 & (1+i)\, k_y \\
 (1+i)\, a & (1-i)\, k_y & a-k_z \\
\end{bmatrix}.
\end{align}
The matrix is NH for $a \neq 0 $. Defining the variables, $\omega = \frac{-1+\sqrt{3} \,i}{2} $ and
$\alpha_\pm  \equiv \sqrt[3]{ q \pm \sqrt{ p^3+ q^2}}$, the 3 eigenvalues turn out to be
\begin{align}
\label{eqPTev}
& \lbrace E_1, \, E_2,\, E_3 \rbrace  = 
\lbrace\alpha_+ +\alpha_-  , \, \omega \, \alpha_+ +  \omega^*  \, \alpha_- ,
\, \omega^* \, \alpha_+ +  \omega  \, \alpha_- \rbrace\,,\nn
& \text{where } p = \frac{ k_x^2+2 \,k_y^2+k_z^2-a^2-2 \,a\, k_z} {3} \text{ and }
q =  \frac{  \left(k_x^2-2\, k_y^2\right)\left(a-k_z\right) } {2} \,.
\end{align}
The real-valued function $p^3 + q^2$ plays a key role as surfaces hosting EP$_2$s appear at $p^3 + q^2 = 0$ (let us label the singular surface as ES$_2$). Additionally, the curve where the surfaces for $p = 0$ and $q =0$ vanish form an EC$_3$, characterised by the Jordan-normal form, $ \mathcal {J}_3 (0) $. Fig.~\ref{figpt} shows the nature of the various surfaces and the knotted EC$_3$ curve for $a=1.25$.

Here, PT-symmetry allows a perturbation of the form of $B_{3,3}$, thus allowing cube-root singularities. 
Instead of trying to write down the complicated expressions of the eigenvalues for such generic perturbations, we focus on two interesting cases when the singularities can be identified easily via simple analytical expressions:
\begin{enumerate}

\item For $k_z= 0$, we have $p=0$ along the ellipse, $k_x^2 + 2\, k_y^2 =  a^2$, and $q= 0 $ along the lines $k_x = \pm \,\sqrt 2 \,k_y$.
Four EP$_3$s appear at $\lbrace k_z = 0, \, k_x = \pm 1/\sqrt 2, \, k_y = \pm a/ 2 \rbrace$. We identify 2 distinct situations: 
\\(a) Adding a diagonal perturbation of the form, $\text{diag} \lbrace \,k_z,\,0,\, -\,k_z \rbrace $ (with $k_z \sim \epsilon $), upon moving out of the $k_z = 0$ plane, the eigenvalues have a Puiseux series starting with the power-law, $k_z^{1/2}$. In particular, we find that the eigenvalues bifurcate as $\lbrace 0, \, \pm \sqrt[6]{-1} \sqrt{k_z \left(k_z-2 a\right)} \rbrace $, for the EP$_3$ at $\lbrace k_z = 0, \, k_x =  1/\sqrt 2, \, k_y =  a/ 2 \rbrace $.
\\(b) Adding a PT-symmetry-preserving perturbation of the form,
 \begin{align}
\epsilon \begin{bmatrix}
 0 & i \,a_2 & \frac{(1-i)\, a_2} {\sqrt{2}} \\
 i \,a_2 & 0 & 0 \\
 \frac{(1+i) \,a_2} {\sqrt{2}} & 0 & 0 \\
 \end{bmatrix},
 \end{align}
with respect to the location of an EP$_3$ (which does not project out of the $k_z = 0$), we have an upper-3 Hessenberg matrix, $B_{3,3}$. It splits the threefold-degenerate zero eigenvalue into $\lbrace  \varepsilon_3  ,\, \pm \,(-1)^{2/3}\,  \varepsilon_3   \rbrace $, where $ \varepsilon_3 =  \sqrt[3]{a \,a_2^2 \,\epsilon -3 \,a^2\, a_2  }\; \epsilon^{1/3}$.
Hence, it produces a leading-order splitting $\sim \epsilon^{1/3}$.

\item For $k_z= a $, we have $q=0$ (identically) and $ p = 0 $ along the ellipse, $k_x^2 + 2\, k_y^2 = 2\, a^2$. Here, we get EC$_3$s along this elliptical curve. Adding a diagonal perturbation of the form, $\epsilon \,\text{diag} \lbrace \,1,\,0,\, -\,1 \rbrace $, upon moving out of the $k_z = a$ plane, the eigenvalues have a Puiseux series starting with the power-law, $ \epsilon^{1/3}$. Needless to say, a generic PT-symmetric $B_{3,3}$-type shows a leading-order splitting of the cube-root type. In particular, adding a PT-symmetry-preserving perturbation of the form,
 \begin{align}
\epsilon \begin{bmatrix}
 0 & 0 & (1-i)\, a_3 \\
 0 & 0 & (1+i)\, a_3 \\
 (1+i)\, a_3 & -(1-i)\,a_3 & 0 \\
 \end{bmatrix},
 \end{align}
with respect to the location of the EC$_3$, the threefold-degenerate zero eigenvalue splits into $\lbrace  \varepsilon_3  ,\,
\pm \,(-1)^{1/3}\,  \varepsilon_3   \rbrace $, where $ \varepsilon_3 =  2^{5/6}\, \sqrt[3]{-a^2\, a_3\, ( 1 + \sin \alpha ) \cos \alpha }
 \;\epsilon^{1/3}$.
 
\end{enumerate}

\section{Discussions and future outlook}
\label{secsum}

In this paper, we have outlined an analysis to determine the order of singularitites of exceptional degeneracies, given an $s$-dimensional complex square matrix. In particular, by determining its symmetry class \cite{bernard}, the kind of allowed symmetry-preserving perturbations (about the singular matrix at an EP) is constrained. On transforming an allowed perturbing matrix to the Jordan-normal basis of the defective matrix, its upper-$\ell$ Hessenberg structure (where $\ell$ is a positive integer $\leq s $) is revealed. It tells us the leading-order algebraic singularity (i.e., a branch point) that characterises the corresponding EP, which also shows up as the leading-order fractional power, $1/\ell$, of the deviation-parameter ($\epsilon$) away from the EP --- the degenerate eigenvalues and eigenvectors split as $\epsilon^{1/\ell} $. After providing the generic prescription, we have moved onto apply the formalism to three-band models, considering three different symmetry classes (viz. P-, C-, and PT-symmetries). We have proven that the most singular behaviour that can appear in the P- and C-symmetric NH Hamiltonians is of the square-root variety. On the other hand, PT-symmetric NH Hamiltonians can generically host the strongest singular behaviour possible for three-band systems, namely $\epsilon^{1/3} $. Because of the spectator flat-band in the three-band P- and C-symmetric systems (constrained to be present by the symmetries), the other two bands are negative of each other, which immediately tells us that there is no possibility getting exceptional singularities of order two. On the other hand, we do get EC$_2$ (singular curves) in the 3D PT-symmetric systems with three bands. Overall, we have demonstrated the emergence of singularities of all possible co-dimensions --- EPs, ECs, and ESs --- in the various systems studied here. Continuing our procedure for higher-band models will tell us the strongest algebraic singularities around their EPs. For example, four-band P- and C-symmetric $4\times 4$ matrices allow EP$_4$s which split as $\epsilon^{1/4}$ and $\epsilon^{1/2}$, respectively (see appendix). The curious behaviour of the the latter originates from the Jordanl-normal comprising the Jordan blocks, $\mathcal{J}_1 (0)$ and $\mathcal{J}_3 (0)$. 

Through our mathematical analysis, we have been able to pinpoint the reason behind the singular behaviour of a particular degree, emerging for an higher-order EP (or EC or ES) in a system protected by a specific symmetry. Furthermore, we have demonstrated that the perturbations can be fine-tuned such that the singular behaviour is endowed with a directional dependence. This has been demonstrated in the (1) P- and C-symmetric systems whose energy-splittings can have either $\sim \epsilon$ (linear) or $\epsilon^{1/2}$ (square-root) nature; and (2) PT-symmetric systems whose energy-splittings can have either $\epsilon^{1/2}$ (square-root) or $\epsilon^{1/3}$ (cube-root) nature. This can be utilised in designing multi-functional EP-based sensors \cite{emil-budich}, which can now have direction-dependent sensitivity. 
We emphasise that the enhanced sensitivity is not fully contingent on exact degeneracy: it originates from the branch-point singularity of the eigenvalues ($ \varepsilon $) in the vicinity of the EP. Since
$ \varepsilon(\epsilon)-\varepsilon(0) \sim \epsilon^{1/\ell} $, a sensor operated at a small but finite $\epsilon$, where the system is nondegenerate, can still benefit from the enhanced spectral response associated with the nearby EP. We note, however, that enhanced spectral sensitivity alone does not necessarily imply enhanced measurement-precision, as the full sensing response is also influenced by eigenvector dynamics and noise properties~\cite{Langbein,emil-budich}.

While our derivation is presented using explicit matrix forms for transparency, the result admits a more general linear-algebraic interpretation. At an EP, the Hamiltonian is brought to its Jordan normal form, whose generalised eigenvectors constitute a Jordan chain. A symmetry operator $S$ does not act independently on the vectors in this chain. Rather, the symmetry relation, $S \,H\, S^{-1}=\eta \,H$, or $S\, H^{T}\, S^{-1}=\eta \,H$, depending on the symmetry class, requires every symmetry-preserving perturbation to belong to the corresponding intertwiner space of the Jordan representation. In the Jordan basis, this intertwining condition is expected to translate into linear constraints on the perturbation matrix elements, which we find to determine the admissible Hessenberg structure in the cases considered. For the $P$- and $C$-symmetry classes considered here, the symmetry operators square to the identity and induce a $\mathbb{Z}_2$-grading of the Hilbert space. The symmetry condition constrains both the Hamiltonian and every symmetry-preserving perturbation to respect this grading, yielding the characteristic block-off-diagonal structure in the natural basis. Upon transforming to the Jordan basis of the defective matrix, we find, consistent with our explicit construction, that these grading-constraints restrict the perturbation to the class of upper-2 Hessenberg matrices.
By contrast, $PT$ symmetry is implemented by the antiunitary constraint $H=(PK)\,H^{*}\,(PK)^{-1}$, which is an antilinear rather than a complex-linear condition on the Hamiltonian. Consequently, it does not induce an analogous $\mathbb{Z}_2$-grading of the complex vector space, and the corresponding intertwining constraints are correspondingly less restrictive. In the Jordan basis, they permit the larger class of upper-3 Hessenberg perturbations obtained in our explicit constructions. We thus expect the Hessenberg bandwidth to be governed by the structure of the symmetry-invariant intertwiner space associated with the Jordan block, rather than by the particular matrix parametrisation used in our calculations. We have verified this correspondence explicitly rather than in full generality.
Finally, we note that our choice of symmetry-operator matrices is one convenient realisation of the corresponding symmetry class. One can check explicitly that any other realisation of the symmetry operators leaves both the Jordan normal form and the symmetry constraints and, hence, the admissible perturbation space and Hessenberg bandwidth unchanged.

Our terminology involving ``topological phases'' and``topologically-robust phenomena'' follows the established fact that nodal degeneracies are far more abundant and stable in NH systems than in their Hermitian counterparts, as reviewed in Ref.~\cite{emil_review}. In a Hermitian system, a generic band-touching requires three real conditions to be satisfied simultaneously --- therefore, such degeneracies are fine-tuned in 2D and become stable only in 3D. In NH systems, by contrast, a degeneracy is fixed by the vanishing of the complex discriminant of the characteristic polynomial, requiring only two real conditions. Therefore, EPs are already generic and stable in 2D, while still carrying a well-defined topological charge associated with spectral winding (vorticity) \cite{Shen2018, Kawabata2019}. Recent work has further established EPs as topological singularities separating distinct NH spectral phases \cite{flore-elisabet, flore-elisabet0, ips-bp}. It is this reduced-codimension stability of EPs and, in the symmetry-constrained cases treated here, their additional protection under various symmetries (as established in our earlier work \cite{ips-emil}) motivate our terminology. We stress that this is a statement about the local stability and charge of the exceptional degeneracy itself, not about a global bulk invariant. A genuine bulk invariant, such as the biorthogonal polarisation, is indeed computable in NH lattice models and has been shown to jump precisely at bulk phase transitions \cite{flore-elisabet, flore-elisabet0, ips-bp}. However, no such invariant is computed or compared across parameter regions in the present work, and we do not claim a conventional bulk topological phase transition. Last but not the least, emergence of EPs do indeed characterise phase transitions in systems hosting chiral Majorana zero modes \cite{ips-ep-epl, ips-tewari}.

In the future, it will be worthwhile to explore how to design a bidirectional sensor, based on our theoretical observations. Another interesting direction is to investigate if and how the behaviour of non-Hermitian skin effects (NHSEs) \cite{kang-emil, ips-yao-lee} change with the degree of the EP-singularity and whether this can be traced back to the upper-$\ell$ value of the perturbing Hessenberg matrix. Attached to this is the question of the directional dependence of the NHSEs connected with the differing eigensystem-splittings depending on the perturbing parameters, whether there are NH topological invariants to characterise the different scenarios \cite{flore-elisabet0, flore-elisabet, maria_emil, ips-bp}, and whether the distinctions can be meaningfully quantified and identified in processes like topological pumping of charges \cite{zhenming, ips-NH-pumping}.

\section*{Acknowledgments}

This research, leading to the results reported, has received funding from the Council of Science \& Technology (CST), U.P.

\appendix

\section{P-symmetric $4 \times 4$ matrix}
\label{app1}

We impose P-symmetry on a $4\times 4$ non-Hermitian matrix, which constrains its form to \cite{ips-emil}
\begin{align}
\mathcal H^{\rm P}_4 = \begin{bmatrix}
 0 & 0 & a & b \\
 0 & 0 & c & d \\
 e & f & 0 & 0 \\
 g & h & 0 & 0 \\
\end{bmatrix}, \quad
P = \text{diag}\lbrace 1,\,1,\,-1,\,-1 \rbrace ,\quad
P\,\mathcal H^{\rm P}_4\, P^{-1} = - \, \mathcal H^{\rm P}_4 \,.
\end{align}
The $4 \times 4 $ matrix $P$ forms a four-dimensional realisation of the P-symmetry. 
For $ a = b\, c/d $ and $ g= -\, (a \,e+c\, f+d\, h)/b $, it hosts an EP$_4$ with the fourfold-degenerate eigenvalue, $\lambda = 0$. At the defective point, its Jordan-normal form is
\begin{align}
S^{-1} \, {\mathcal H^{\rm P}_{\rm EP}} \, S = \mathcal J_4(0)\,, \text{ with }
S = \begin{bmatrix}
0 & -\frac{ b\, d}{b \,c\, e+c\, d\, f} & 0 & -\frac{ b\, d^3 h}{c^2 \, (b \,e+d\, f)^2 \, (c\, f+d \,h)} \\
 0 & -\frac{d^2}{b \,c\, e+c\, d\, f} & 0 & -\frac{d^2 (b \,c\, e +d \, (c\, f+d \,h))}{c^2 \, (b \,e+d\, f)^2 \, (c\, f+d \,h)} \\
 -\frac{d}{c} & 0 & -\frac{d^2}{c^2 \, (b \,e+d\, f)} & 0 \\
 1 & 0 & 0 & 0
\end{bmatrix} 
\text{ and }
{\mathcal H^{\rm P}_{\rm EP}}  = {\mathcal H^{\rm P}_4} \Big \vert_{a = \frac{b\, c} {d},\; g= \frac{a \,e+c\, f+d\, h}{ -\,b}}.
\end{align}
Now let us add a P-symmetric perturbation, $\epsilon\, \tilde B^{\rm P}_{4,4}$, to ${\mathcal H^{\rm P}_{\rm EP}}$, with
\begin{align}
\tilde B^{\rm P}_{4,4} = \begin{bmatrix}
0 & 0 & a_1 & a_2 \\
 0 & 0 & a_3 & a_4 \\
 a_5 & a_6 & 0 & 0 \\
 a_7 & a_8 & 0 & 0 \\
\end{bmatrix}.
\end{align}
Explicit calculation shows that $S^{-1}\, \tilde B^{\rm P}_{4,4}\, S $ is an upper-4 Hessenberg matrix, $ B_{4,4}$, which we do not write here explicitly due to its lengthy expression. All the four eigenvalues behave as $ \epsilon^{1/4} $, which we do not show here due to their complicated analytical forms.

\section{C-symmetric $4 \times 4$ matrix}
\label{app2}

We impose C-symmetry on a $4\times 4$ non-Hermitian matrix, such that it gets constrained to the form,
\begin{align}
\label{eqcdim4}
\mathcal H^{\rm C}_4 = \begin{bmatrix}
a & b & c & 0 \\
 d & e & 0 & c \\
 g & 0 & -e & b \\
 0 & g & d & -a \\
\end{bmatrix}, \quad
C =  \begin{bmatrix}
 0 & 0 & 0 & -1 \\
 0 & 0 & 1 & 0 \\
 0 & 1 & 0 & 0 \\
 -1 & 0 & 0 & 0 \\
\end{bmatrix},\quad
C \left( \mathcal H^{\rm C}_4 \right)^T C^{-1} = \epsilon_c  \, \mathcal H^{\rm C}_4 \text{ with } \epsilon_c = -1 \,.
\end{align}
The $4 \times 4 $ matrix $C$ forms a four-dimensional realisation of the C-symmetry with $\epsilon_c=-1$. 
For $ c = -\,(a+e)^2 /(4\, g) $ and $ d= -\,(a^2+2 \,c\, g+e^2)/(2 \,b)$, it hosts an EP$_4$ with the fourfold-degenerate eigenvalue, $\lambda = 0$. At the defective point, its Jordan-normal form is
\begin{align}
& S^{-1} \, {\mathcal H^{\rm C}_{\rm EP}} \, S =  \begin{bmatrix}
\mathcal J_1(0) & 0 \\
 0 & \mathcal J_3(0) \\
\end{bmatrix}, \text{ with }
S = \begin{bmatrix}
 0 & -\frac{b (a+e)^2}{2 g} & 0 & 0 \\
 \frac{(a+e)^2}{8 b g} & \frac{(a-e) (a+e)^2}{4 g} & -\frac{(a+e)^2}{4 g} & 0 \\
 \frac{1}{2} & -b (a+e) & b & 0 \\
 \frac{e}{2 b} & \frac{1}{2} (a-e) (a+e) & -a & 1 \\
\end{bmatrix} \nn
& \text{and }
{\mathcal H^{\rm C}_{\rm EP}}  = {\mathcal H^{\rm C}_4} 
\Big \vert_{c = -\frac{(a+e)^2}{4\, g},\; d = -\frac{a^2+2 \,c\, g+e^2}{2 \,b} }.
\end{align}
Intriguingly, the Jordan-normal form is split into two Jordan blocks, which needs special care. In the Jordan normal basis, we add a perturbation of the form
\begin{align}
\mathcal B = \epsilon \begin{bmatrix}
 0 & 0 \\
0  & B_{3,2}\\
\end{bmatrix}, \text{ setting }
B_{3,2} = \begin{bmatrix}
 a_1 & 0 & 0 \\
 1 & 0 & 0 \\
0 & 1 & -a_1 \\
 \end{bmatrix},
\end{align}
where $B_{3,2}$ is a three-dimensional upper-2 Hessenberg matrix.
This preserves the PT-symmetric form shown in Eq.~\eqref{eqcdim4}, because
\begin{align}
S^{-1} {\mathcal B} \,S = \epsilon \begin{bmatrix}
  a_1 & 0 & 0 & 0 \\
 \frac{a_1 (e-a)+1} {2\, b} & 0 & 0 & 0 \\
 \frac{2 \,g \left(a_1 (a+e)-1\right)}{(a+e)^2} & 0 & 0 & 0 \\
 0 & \frac{2 \,g \left(a_1 (a+e)-1\right)}{(a+e)^2} & \frac{a_1 (e-a)+1}{2 \,b} & -a_1 \\
 \end{bmatrix}.
\end{align}
Replacing $B_{3,2}$ by a $B_{3,3}$ matrix does not preserve the C-symmetric form and, hence, is not allowed. This simple example tells us that the eigensystem-splitting away from the EP$_4$ should be the same as the C-symmetric three-band models, namely $\propto \epsilon^{1/2}$.
Therefore, adding a C-symmetric perturbation with generic parameters, for example
\begin{align}
\epsilon \begin{bmatrix}
 a_1 & a_2 & a_3 & 0 \\
 a_4 & a_5 & 0 & a_3 \\
 a_6 & 0 & -a_5 & a_2 \\
 0 & a_6 & a_4 & -a_1 \\
\end{bmatrix},
\end{align}
produces eigenvalues behaving as $ \epsilon^{1/2} $ (rather than $\epsilon^{1/4}$). We do not show here the closed-form analytical forms of the eigenvalues (appearing as $\pm $ pairs due to the C-symmetry) because of their lengthy expressions.

\bibliography{ref_nh}

\end{document}